\documentclass[twocolumn,showpacs]{revtex4}
\usepackage{amssymb}
\usepackage{makeidx}
\usepackage{amsmath}
\usepackage{graphicx}
\usepackage{dcolumn}
\usepackage{bm}
\usepackage{epstopdf}

\begin{document}

\title{Nonlinear absorption of high-intensity shortwave radiation in plasma
within relativistic quantum theory }
\author{H.K. Avetissian$^{1}$}
\author{A.G. Ghazaryan$^{1}$}
\author{H.H. Matevosyan$^{2}$}
\author{G.F. Mkrtchian$^{1}$}
\affiliation{$^{1}$ Centre of Strong Fields Physics, Yerevan State University, 1 A.
Manukian, Yerevan 0025, Armenia}
\affiliation{$^{2}$ Plasma Theory Group, Institute of Radiophysics and Electronics, NAS
RA, 0203 Ashtarak, Armenia}
\date{\today }

\begin{abstract}
On the base of the quantum kinetic equation for density matrix in plasma at
the stimulated bremsstrahlung of electrons on ions, the nonlinear absorption
rate for high-intensity shortwave radiation in plasma has been obtained within
relativistic quantum theory. Both classical Maxwellian and degenerate quantum
plasma are considered for x-ray lasers of high intensities. Essentially
different dependences of nonlinear absorption rate on polarization of strong
laser radiation is stated.
\end{abstract}

\maketitle



\section{ Introduction}

At present, due to the rapid advance in free electron laser (FEL)\
technology \cite{FELX1,FELX2,FELX3} became real the implementation of
ultrashort laser pulses in x-ray domain exceeding the intensities $10^{20}\ 
\mathrm{W/cm}^{2}$. For such intensities the electron-wave field energy
exchange over a x-ray wavelength is larger than the photon energy, and the
laser-matter interaction has essentially multiphoton character \cite{Book}.
Thus, in recent decade the wide research field is opened up, where common
nonlinear effects are extended to high energy transitions for various
systems - from atoms \cite{XA1,XA2,XA3,XA4,XA5}, through molecules \cite%
{XM1,XM2,XM3,XM4}, to plasma and solid samples \cite%
{XPS1,XPS2,XPS3,XPS4,XPS5,XPS6}. Among the fundamental processes of
laser-plasma interaction, the inverse bremsstrahlung absorption of an
intense laser field in plasma is one of the contemporary problems that have
applications ranging from plasma diagnostics to thermonuclear reactions and
generation of intense x-ray radiation. Note that in the field of intense
x-ray laser, an electron may gain considerable energies absorbing even of a
few quanta, which makes it as an effective mechanism for laser-plasma
heating. The theoretical description of this phenomenon in such superstrong
radiation fields requires one to go beyond the scope of common Quantum
Electrodynamics -Feynman diagrams corresponding to the perturbation theory.

With the advent of lasers many pioneering papers have been devoted to the
theoretical investigation of the electron-ion scattering processes in gas or
plasma in the presence of a laser field using nonrelativistic \cite%
{CSB1,QSB1,CSB2,Bornnr1,LF1,CSB3,CBS33,CBS333,Bunk,Eik1,LF2,QSB2} as well as
relativistic \cite{Fedor,BornRel1,RLF1, RLF2,RLF4} considerations. The
appearance of superpower ultrashort laser pulses of relativistic intensities
has triggered new interest in stimulated bremsstrahlung (SB) in relativistic
domain, where investigations were carried out in the Born \cite%
{BornRel2,BornRel3,BornRel4}, eikonal \cite{RLF1}, and generalized eikonal 
\cite{GEA} approximations over the scattering potential. Beyond this
approximation for the infrared and optical lasers, in the multiphoton
interaction regime, one can apply classical theory and the main
approximation in the classical theory is low frequency or impact
approximation \cite{CSB3,Bunk}. Low frequency approximation have been
generalized for relativistic case in Refs. \cite{RLF2,RLF4}, where the
effect of an intense EM wave on the dynamics of SB and nonlinear absorption
of intense laser radiation by a monochromatic electron beam and by
relativistic plasma due to the SB have been carried out. For the infrared
and optical lasers the quantum effects in the SB process is smeared out due
to smallness of the photon energy. Meanwhile for a intense x-ray radiation
the quantum nonlinear over the field effects will be considerable. Besides,
relativistic quantum effects may be essential for plasmas of high densities 
\cite{PRE1,PRE2}. Regarding the solid densities, absorption coefficient may
reach considerably large values. As was shown in Ref. \cite{REL}, where
interaction of superstrong lasers with thin plasma targets of solid
densities was investigated via particle-in-cell simulations, the
inverse-bremsstrahlung absorption is dominant for electron densities above $%
10^{21}\ \mathrm{cm}^{-3}$. Thus, it is of interest to consider x-ray
multiphoton absorption via inverse bremsstrahlung in ultradense classical as
well as quantum plasmas.

In the present paper the inverse-bremsstrahlung absorption of an intense
x-ray laser field in the dense classical and quantum plasmas is considered
in the relativistic-quantum regime considering a wave-field exactly, while a
scattering potential of a plasma ions as a perturbation. The x-ray radiation
power absorbed in plasma is investigated for a circularly and linearly
polarized waves (CPW and LPW, respectivelly) arising from the second
quantized consideration. The Liouville-von Neumann equation for the density
matrix is solved analytically for a grand canonical ensemble. With the help
of this solution we investigate the nonlinear inverse-bremsstrahlung
absorption rate for Maxwellian as well as for degenerate quantum plasmas. It
is shown that depending on the intensity of an incident x-ray, one can
achieve the quite large absorption coefficients. Hence the considered
mechanism may serve as an efficient tool for ultrafast plasma heating.

The organization of the paper is as follows. In section II, the relativistic
quantum dynamics of SB is presented with analytical results for density
matrix and inverse-bremsstrahlung absorption rate. In section III, we
consider the problem numerically. Finally, conclusions are given in Section
IV.

\section{Basic Model and Theory}

Let us consider the relativistic quantum theory of plasma nonlinear
interaction with the arbitrary strong EM wave field by microscopic theory of
electrons-ions interaction on the base of a single particle density matrix.
The EM wave with the four-wave vector $k\equiv (\omega /c,\mathbf{k})$ is
described by the four-vector potential%
\begin{equation}
A^{\mu }=(0,\mathbf{A}),  \label{Amju}
\end{equation}%
where $\mathbf{A}$ is defined as:%
\begin{equation}
\mathbf{A}(\tau )=A_{0}\{\mathbf{e}_{1}\cos \omega \tau +\mathbf{e}_{2}g\sin
\omega \tau \};  \label{Avec}
\end{equation}%
\begin{equation*}
\tau =t-\frac{\mathbf{\nu }_{0}\mathbf{r}}{c};\quad \mathbf{\nu }_{0}=\frac{c%
\mathbf{k}}{\omega };\quad \mathbf{e}_{1}\mathbf{\nu }_{0}=\mathbf{e}_{2}%
\mathbf{\nu }_{0}=\mathbf{e}_{1}\mathbf{e}_{2}=0,
\end{equation*}%
with the amplitude $A_{0}$, unit polarization vectors $\mathbf{e}_{1,2}$ and
ellipticity parameter $g$. The ions are assumed to be at rest and being
either randomly or nonrandomly distributed in plasma, the static potential
field of which (for nucleus/ion -as a scattering center- the recoil momentum
is neglected) is described by the scalar potential 
\begin{equation*}
A^{(e)}\left( x\right) =\left( \varphi \left( \mathbf{r}\right) ,0\right) ,
\end{equation*}%
where 
\begin{equation*}
\varphi \left( \mathbf{r}\right) =\sum_{i}^{N_{i}}\varphi _{i}\left( \mathbf{%
r-R}_{i}\right) .
\end{equation*}%
Here $\varphi _{i}$ is the potential of a single ion situated at the
position $\mathbf{R}_{i}$, and $N_{i}$ is the number of ions in the
interaction region.

To investigate the quantum dynamics of SB we need the quantum kinetic
equations for a single particle density matrix, which can be derived arising
from the second quantized formalism. As is known, the Dirac equation allows
the exact solution in the field of a plane EM wave (Volkov solution).
Although the Volkov states are not stationary, as there are no real
transitions in the monochromatic EM wave field (due to the violation of
energy and momentum conservation laws), the state of an electron with a
charge $e$ and mass $m$ in an EM wave field can be characterized by the
quasimomentum $\mathbf{\Pi }$ and polarization $\sigma $, and the particle
state in the field (\ref{Amju}) is given by the wave function:%
\begin{equation*}
\Psi _{\mathbf{\Pi }\sigma }=\left[ 1+\frac{e\left( \gamma k\right) \left(
\gamma A\right) }{2c(kp)}\right] \frac{u_{\sigma }(p)}{\sqrt{2E_{\mathbf{\Pi 
}}\mathcal{V}}}\exp \left[ -\frac{i}{\hbar }\Pi x\right]
\end{equation*}%
\begin{equation*}
\times \exp \left\{ \frac{i}{\hbar }\left[ \frac{eA_{0}}{c(pk)}(\mathbf{e}%
_{1}\mathbf{p}\sin \omega \tau -g\mathbf{e}_{2}\mathbf{p}\cos \omega \tau
)\right. \right.
\end{equation*}%
\begin{equation}
\left. \left. -\frac{e^{2}A_{0}^{2}}{8c^{2}(pk)}(1-g^{2})\sin (2\omega \tau )%
\right] \right\} ,  \label{1.86q}
\end{equation}%
where $\mathcal{V}$ is the quantization volume, $u_{\sigma }$ is the
bispinor amplitude of a free Dirac particle with polarization $\sigma $, and 
$\Pi =(E_{\mathbf{\Pi }}/c,\mathbf{\Pi )}$ is the average four-kinetic
momentum or \textquotedblright quasimomentum\textquotedblright\ of the
particle in the periodic field, which is determined via a free particle
four-momentum $p=(\mathcal{E}/c,\mathbf{p})$ and relativistic invariant
parameter of the wave intensity $\xi _{0}=eA_{0}/mc^{2}$ by the following
equation%
\begin{equation}
\Pi =p+k\frac{m^{2}c^{2}}{4kp}(1+g^{2})\xi _{0}^{2}.  \label{1.87}
\end{equation}%
From this equation follows that%
\begin{equation}
\Pi ^{2}=m^{\ast 2}c^{2};\ m^{\ast }=m\left( 1+\frac{1+g^{2}}{2}\xi
_{0}^{2}\right) ^{1/2},  \label{1.88}
\end{equation}%
where $m^{\ast }$ is the effective mass of the particle in the monochromatic
wave. The states (\ref{1.86q}) are normalized by the condition%
\begin{equation*}
\int \Psi _{\mathbf{\Pi }^{\prime }\sigma ^{\prime }}^{\dagger }\Psi _{%
\mathbf{\Pi }\sigma }d\mathbf{r}=\frac{\left( 2\pi \hbar \right) ^{3}}{%
\mathcal{V}}\delta (\mathbf{\Pi }-\mathbf{\Pi }^{\prime })\delta _{\sigma
,\sigma ^{\prime }}.
\end{equation*}%
Cast in the second quantization formalism, the Hamiltonian is%
\begin{equation}
\mathcal{H}=\int \widehat{\Psi }^{+}\widehat{H}_{0}\widehat{\Psi }d\mathbf{r+%
}\mathcal{H}_{sb},  \label{8.1.5}
\end{equation}%
where $\widehat{\Psi }$ is the fermionic field operator, $\widehat{H}_{0}$
is the one-particle Dirac Hamiltonian in the plane EM wave (\ref{Amju}), and
the interaction Hamiltonian is%
\begin{equation}
\mathcal{H}_{sb}=\frac{1}{c}\int \widehat{j}A^{(e)}d\mathbf{r,}
\label{8.1.6}
\end{equation}%
with the current density operator%
\begin{equation}
\widehat{j}=e\widehat{\Psi }^{+}\gamma _{0}\gamma \widehat{\Psi }.
\label{8.1.7}
\end{equation}%
Making Fourier transformation%
\begin{equation*}
A^{(e)}(x)=\frac{1}{\left( 2\pi \right) ^{3}}\int A^{(e)}\left( \mathbf{q}%
\right) e^{-i\mathbf{qr}}d\mathbf{q},
\end{equation*}%
the expression (\ref{8.1.6}) will have a form%
\begin{equation}
\mathcal{H}_{sb}=\frac{1}{c\left( 2\pi \right) ^{3}}\int \widehat{\Psi }%
^{+}V\left( \mathbf{q}\right) e^{-i\mathbf{qr}}\widehat{\Psi }d\mathbf{q}d%
\mathbf{r},  \label{1.109}
\end{equation}%
where 
\begin{equation}
V\left( \mathbf{q}\right) =\int \sum_{i}^{N_{i}}e\varphi _{i}\left( \mathbf{%
r-R}_{i}\right) e^{-i\mathbf{qr}}d\mathbf{r.}  \label{Vq}
\end{equation}

We pass to the Furry representation and write the Heisenberg field operator
of the electron in the form of an expansion in the quasistationary Volkov
states (\ref{1.86q})%
\begin{equation}
\widehat{\Psi }(\mathbf{r},t)=\sum_{\sigma }\int d\Phi _{\mathbf{\Pi }}%
\widehat{a}_{\mathbf{\Pi },\sigma }e^{\frac{i}{\hbar }E_{\mathbf{\Pi }%
}t}\Psi _{\mathbf{\Pi }\sigma }(\mathbf{r},t),  \label{8.1.8}
\end{equation}%
where $d\Phi _{\mathbf{\Pi }}=\mathcal{V}d^{3}\mathbf{\Pi }/\left( 2\pi
\hbar \right) ^{3}$. In Eq. (\ref{8.1.8}) we have excluded the antiparticle
operators, since contribution of electron-positron intermediate states will
be negligible for considered intensities and photon energies $\varepsilon
_{\gamma }=\hbar \omega \ll mc^{2}$. The creation and annihilation
operators, $\widehat{a}_{\mathbf{\Pi },\sigma }^{+}(t)$ and $\widehat{a}_{%
\mathbf{\Pi },\sigma }(t)$, associated with positive energy solutions
satisfy the anticommutation rules at equal times%
\begin{equation}
\{\widehat{a}_{\mathbf{\Pi },\sigma }^{\dagger }(t),\widehat{a}_{\mathbf{\Pi 
}^{\prime },\sigma ^{\prime }}(t^{\prime })\}_{t=t^{\prime }}=\frac{\left(
2\pi \hbar \right) ^{3}}{\mathcal{V}}\delta \left( \mathbf{\Pi -\Pi }%
^{\prime }\right) \delta _{\sigma ,\sigma ^{\prime }}\;,\;  \label{8.1.9a}
\end{equation}%
\begin{equation}
\{\widehat{a}_{\mathbf{\Pi },\sigma }^{\dagger }(t),\widehat{a}_{\mathbf{\Pi 
}^{\prime },\sigma ^{\prime }}^{\dagger }(t^{\prime })\}_{t=t^{\prime }}=\{%
\widehat{a}_{\mathbf{\Pi },\sigma }(t),\widehat{a}_{\mathbf{\Pi }^{\prime
},\sigma ^{\prime }}(t^{\prime })\}_{t=t^{\prime }}=0.  \label{8.1.9b}
\end{equation}%
Taking into account Eqs. (\ref{8.1.8}), (\ref{8.1.7}), (\ref{8.1.6}) and (%
\ref{1.86q}), the second quantized Hamiltonian can be expressed in the form%
\begin{equation}
\mathcal{H}=\mathcal{H}_{0}+\mathcal{H}_{sb}\left( t\right) .  \label{SH}
\end{equation}%
The first term in Eq. (\ref{SH}) is the Hamiltonian of Volkov dressed
electron field%
\begin{equation}
\mathcal{H}_{0}=\sum_{\sigma }\int d\Phi _{\mathbf{\Pi }}E_{\mathbf{\Pi }}%
\widehat{a}_{\mathbf{\Pi },\sigma }^{+}\widehat{a}_{\mathbf{\Pi },\sigma },
\label{V}
\end{equation}%
while the second term 
\begin{equation}
\mathcal{H}_{sb}\left( t\right) =\sum_{\sigma \sigma ^{\prime }}\int d\Phi _{%
\mathbf{\Pi }}\int d\Phi _{\mathbf{\Pi }^{\prime }}M_{\mathbf{\Pi }^{\prime
},\sigma ^{\prime };\mathbf{\Pi },\sigma }\left( t\right) \widehat{a}_{%
\mathbf{\Pi }^{\prime },\sigma ^{\prime }}^{+}\widehat{a}_{\mathbf{\Pi }%
,\sigma }  \label{IBS}
\end{equation}%
is the interaction Hamiltonian describing the SB with amplitudes%
\begin{equation}
M_{\mathbf{\Pi }^{\prime },\sigma ^{\prime };\mathbf{\Pi },\sigma }\left(
t\right) =\frac{1}{\mathcal{V}}\sum\limits_{s=-\infty }^{\infty
}e^{-is\omega t}\mathcal{M}_{\mathbf{\Pi }^{\prime },\sigma ^{\prime };%
\mathbf{\Pi },\sigma }^{(s)},  \label{part}
\end{equation}%
\begin{equation*}
\mathcal{M}_{\mathbf{\Pi }^{\prime },\sigma ^{\prime };\mathbf{\Pi },\sigma
}^{(s)}=\frac{V\left( \mathbf{q}_{s}\right) }{2c\sqrt{E_{\mathbf{\Pi }}E_{%
\mathbf{\Pi }^{\prime }}}}\overline{u}_{\sigma ^{\prime }}(p^{\prime })
\end{equation*}%
\begin{equation*}
\times \left[ \widehat{\epsilon }_{0}B_{s}+\left( \frac{e\widehat{B}_{1s}%
\widehat{k}\widehat{\epsilon }_{0}}{2c(kp^{\prime })}+\frac{e\widehat{%
\epsilon }_{0}\widehat{k}\widehat{B}_{1s}}{2c(kp)}\right) \right.
\end{equation*}%
\begin{equation}
\left. +\frac{e^{2}(k\epsilon _{0})B_{2s}}{2c^{2}(kp^{\prime })(kp)}\widehat{%
k}\right] u_{\sigma }(p).  \label{M}
\end{equation}

In Eq. (\ref{M}) the vector functions $B_{1s}^{\mu }=\left( 0,\mathbf{B}%
_{1s}\right) $ and scalar functions $B_{s}$, $B_{2s}$ are expressed via the
generalized Bessel functions $G_{s}(\alpha ,\beta ,\varphi )$:%
\begin{equation}
G_{s}(\alpha ,\beta ,\varphi )=\sum\limits_{k=-\infty }^{\infty
}J_{2k-s}(\alpha )J_{k}(\beta )e^{i\left( s-2k\right) \varphi },
\label{1.52}
\end{equation}%
\begin{equation*}
\mathbf{B}_{1s}=\frac{A_{0}}{2}\left\{ \mathbf{e}_{1}\left( G_{s-1}(\alpha
,\beta ,\varphi )+G_{s+1}(\alpha ,\beta ,\varphi )\right) \right.
\end{equation*}%
\begin{equation}
\left. \mathbf{+}i\mathbf{e}_{2}g\left( G_{s-1}(\alpha ,\beta ,\varphi
)-G_{s+1}(\alpha ,\beta ,\varphi )\right) \right\} ,  \label{1.111a}
\end{equation}%
\begin{equation}
B_{s}=G_{s}(\alpha ,\beta ,\varphi ),  \label{1.111b}
\end{equation}%
\begin{equation*}
B_{2s}=\frac{A_{0}^{2}}{2}(1+g^{2})G_{s}+\frac{A_{0}^{2}}{2}(1-g^{2})
\end{equation*}%
\begin{equation}
\times \left( G_{s-2}(\alpha ,\beta ,\varphi )+G_{s+2}(\alpha ,\beta
,\varphi )\right) ,  \label{1.111c}
\end{equation}%
and%
\begin{equation}
\hbar \mathbf{q}_{s}\mathbf{=\Pi }^{\prime }-\mathbf{\Pi }-s\hbar \mathbf{k}
\label{1.111d}
\end{equation}%
is the recoil momentum. The definition of the arguments $\alpha ,\beta
,\varphi $ are:%
\begin{equation}
\alpha =\frac{eA_{0}}{\hbar c}\left[ \left( \frac{\mathbf{e}_{1}\mathbf{p}}{%
pk}-\frac{\mathbf{e}_{1}\mathbf{p}^{\prime }}{p^{\prime }k}\right)
^{2}+g^{2}\left( \frac{\mathbf{e}_{2}\mathbf{p}}{pk}-\frac{\mathbf{e}_{2}%
\mathbf{p}^{\prime }}{p^{\prime }k}\right) ^{2}\right] ^{1/2},  \label{1.93a}
\end{equation}%
\begin{equation}
\beta =\frac{e^{2}A_{0}^{2}}{8\hbar c^{2}}(1-g^{2})\left( \frac{1}{pk}-\frac{%
1}{p^{\prime }k}\right) ,  \label{1.93b}
\end{equation}%
\begin{equation}
\tan \varphi =\frac{g\left( \frac{\mathbf{e}_{2}\mathbf{p}}{pk}-\frac{%
\mathbf{e}_{2}\mathbf{p}^{\prime }}{p^{\prime }k}\right) }{\left( \frac{%
\mathbf{e}_{1}\mathbf{p}}{pk}-\frac{\mathbf{e}_{1}\mathbf{p}^{\prime }}{%
p^{\prime }k}\right) }.  \label{1.93c}
\end{equation}

Thus, in order to develop microscopic relativistic quantum theory of the
multiphoton inverse-bremsstrahlung absorption of ultrastrong shortwave laser
radiation in plasma we need to solve the Liouville-von Neumann equation for
the density matrix $\widehat{\rho }$:%
\begin{equation}
\frac{\partial \widehat{\rho }}{\partial t}=\frac{i}{\hbar }\left[ \widehat{%
\rho },\mathcal{H}_{0}+\mathcal{H}_{sb}\left( t\right) \right] ,
\label{EvEq}
\end{equation}%
with the initial condition 
\begin{equation}
\widehat{\rho }\left( -\infty \right) =\widehat{\rho }_{G}.  \label{incon}
\end{equation}%
Here $\widehat{\rho }_{G}$ is the density matrix of the grand canonical
ensemble:%
\begin{equation}
\widehat{\rho }_{G}=\exp \left[ \frac{1}{T_{e}}\left( \Omega +\sum_{\sigma
}\int d\Phi _{\mathbf{\Pi }}\left( \mu -E_{\mathbf{\Pi }}\right) \widehat{a}%
_{\mathbf{\Pi },\sigma }^{+}\widehat{a}_{\mathbf{\Pi },\sigma }\right) %
\right] .  \label{dm}
\end{equation}%
In Eq. (\ref{dm}) $T_{e}$ is the electrons temperature in energy units, $\mu 
$ is the chemical potential, and $\Omega $ is the grand potential. Note that
the initial one-particle density matrix in momentum space is%
\begin{equation*}
\rho _{\sigma _{1}\sigma _{2}}(\mathbf{\Pi }_{1},\mathbf{\Pi }_{2},-\infty )=%
\mathrm{Tr}\left( \widehat{\rho }_{G}\widehat{a}_{\mathbf{\Pi }_{2},\sigma
_{2}}^{+}\widehat{a}_{\mathbf{\Pi }_{1},\sigma _{1}}\right)
\end{equation*}%
\begin{equation}
=n\left( E_{\mathbf{\Pi }_{1}}\right) \frac{\left( 2\pi \hbar \right) ^{3}}{%
\mathcal{V}}\delta \left( \mathbf{\Pi }_{1}\mathbf{-\Pi }_{2}\right) \delta
_{\sigma _{1},\sigma _{2}},  \label{FD1}
\end{equation}%
where%
\begin{equation}
n\left( E_{\mathbf{\Pi }_{1}}\right) =\frac{1}{\exp \left[ \frac{E_{\mathbf{%
\Pi }_{1}}-\mu }{T_{e}}\right] +1}.  \label{FD2}
\end{equation}

We consider Volkov dressed SB Hamiltonian $\mathcal{H}_{sb}\left( t\right) $
as a perturbation. Accordingly, we expand the density matrix as 
\begin{equation*}
\widehat{\rho }=\widehat{\rho }_{G}+\widehat{\rho }^{(1)}.
\end{equation*}%
Then taking into account the relations 
\begin{equation*}
\left[ \widehat{a}_{\mathbf{\Pi }^{\prime },\sigma ^{\prime }}^{+}\widehat{a}%
_{\mathbf{\Pi },\sigma },\widehat{\rho }_{G}\right] =\left( 1-e^{\frac{1}{%
T_{e}}\left( E_{\mathbf{\Pi }^{\prime }}-E_{\mathbf{\Pi }}\right) }\right) 
\widehat{\rho }_{G}\widehat{a}_{\mathbf{\Pi }^{\prime },\sigma ^{\prime
}}^{+}\widehat{a}_{\mathbf{\Pi },\sigma }^{\prime }
\end{equation*}%
and 
\begin{equation*}
\left[ \widehat{\rho }_{G},\mathcal{H}_{0}\right] =0,
\end{equation*}%
for $\widehat{\rho }^{(1)}$ we obtain 
\begin{equation*}
\widehat{\rho }^{(1)}=\frac{1}{i\hbar }\int_{-\infty }^{t}dt^{\prime
}\sum_{\sigma \sigma ^{\prime }}\int d\Phi _{\mathbf{\Pi }}\int d\Phi _{%
\mathbf{\Pi }^{\prime }}M_{\mathbf{\Pi }^{\prime },\sigma ^{\prime };\mathbf{%
\Pi },\sigma }\left( t^{\prime }\right)
\end{equation*}%
\begin{equation}
\times e^{\frac{i}{\hbar }\left( t^{\prime }-t\right) \left( E_{\mathbf{\Pi }%
^{\prime }}-E_{\mathbf{\Pi }}\right) }\left( 1-e^{\frac{1}{T_{e}}\left( E_{%
\mathbf{\Pi }^{\prime }}-E_{\mathbf{\Pi }}\right) }\right) \widehat{\rho }%
_{G}\widehat{a}_{\mathbf{\Pi }^{\prime },\sigma ^{\prime }}^{+}\widehat{a}_{%
\mathbf{\Pi },\sigma }.  \label{sol}
\end{equation}%
Now with the help of this solution one can calculate the desired physical
characteristics of the SB process. In particular, for the energy absorption
rate by the electrons due to the inverse stimulated bremsstrahlung one can
write%
\begin{equation}
\frac{d\mathcal{E}}{dt}=\mathrm{Tr}\left( \widehat{\rho }^{(1)}\frac{%
\partial \mathcal{H}_{sb}\left( t\right) }{\partial t}\right) .  \label{de}
\end{equation}%
It is more convenient to represent the rate of the inverse-bremsstrahlung
absorption via the mean number of absorbed photons by per electron, per unit
time:%
\begin{equation}
\frac{dN_{\gamma e}}{dt}=\frac{1}{\hbar \omega N_{e}}\frac{d\mathcal{E}}{dt},
\label{dn}
\end{equation}%
where $N_{e}$ is the number of electrons in the interaction region. Taking
into account decomposition%
\begin{equation*}
\left( 1-e^{\frac{1}{T_{e}}\left( E_{\mathbf{1}}-E_{\mathbf{2}}\right)
}\right) \mathrm{Tr}\left( \widehat{\rho }_{G}\widehat{a}_{\mathbf{1}}^{+}%
\widehat{a}_{\mathbf{2}}\widehat{a}_{\mathbf{3}}^{+}\widehat{a}_{\mathbf{4}%
}\right)
\end{equation*}%
\begin{equation*}
=\left( 1-e^{\frac{1}{T_{e}}\left( E_{\mathbf{1}}-E_{\mathbf{2}}\right)
}\right) n_{1}\left( 1-n_{2}\right) \delta _{23}\delta _{14},
\end{equation*}%
with the help of Eqs. (\ref{sol}), (\ref{de}), (\ref{dn}), and (\ref{M}) for
large $t$ we obtain 
\begin{equation}
\frac{dN_{\gamma e}}{dt}=\sum_{s=1}^{\infty }\frac{dN_{\gamma e}\left(
s\right) }{dt},  \label{rate}
\end{equation}%
where the partial $s$-photon absorption rates are given by the formula 
\begin{equation*}
\frac{dN_{\gamma e}\left( s\right) }{dt}=\frac{8\pi s}{\hbar N_{e}\mathcal{V}%
^{2}}\int \int \frac{d\Phi _{\mathbf{\Pi }}d\Phi _{\mathbf{\Pi }^{\prime }}}{%
E_{\mathbf{\Pi }}E_{\mathbf{\Pi }^{\prime }}}\left\vert V\left( \mathbf{q}%
_{s}\right) \right\vert ^{2}\left\vert \mathcal{B}_{\mathbf{\Pi }^{\prime };%
\mathbf{\Pi }}^{(s)}\right\vert ^{2}
\end{equation*}%
\begin{equation*}
\times \delta \left( E_{\mathbf{\Pi }^{\prime }}-E_{\mathbf{\Pi }}+s\hbar
\omega \right) \left( 1-e^{\frac{1}{T_{e}}\left( E_{\mathbf{\Pi }^{\prime
}}-E_{\mathbf{\Pi }}\right) }\right)
\end{equation*}%
\begin{equation}
\times n\left( E_{\mathbf{\Pi }^{\prime }}\right) \left( 1-n\left( E_{%
\mathbf{\Pi }}\right) \right) ,  \label{Prate}
\end{equation}%
where%
\begin{equation*}
\left\vert \mathcal{B}_{\mathbf{\Pi }^{\prime };\mathbf{\Pi }%
}^{(s)}\right\vert ^{2}=\left\{ \left\vert \mathcal{E}B_{s}-\frac{e\left( 
\mathbf{pB}_{1s}\right) \omega }{\left( kp\right) c}+\frac{e^{2}\omega B_{2s}%
}{2c^{2}(kp)}\right\vert ^{2}\right.
\end{equation*}%
\begin{subequations}
\begin{equation}
\left. -\frac{\hbar ^{2}\mathbf{q}_{s}^{2}c^{2}}{4}\left\vert
B_{s}\right\vert ^{2}+\frac{e^{2}\hbar ^{2}\left[ \mathbf{kq}_{s}\right] ^{2}%
}{4(kp^{\prime })(kp)}\left[ \left\vert \mathbf{B}_{1s}\right\vert ^{2}-%
\mathrm{Re}\left( B_{2s}B_{s}^{\ast }\right) \right] \right\} ,  \label{Bss}
\end{equation}%
and $\delta (x)$ is the Dirac delta function that expresses the energy
conservation law in SB process. The obtained expression for the absorption
rate is general and applicable to arbitrary polarization, frequency and
intensity of the wave-field. This formula is applicable for a grand
canonical ensemble and is always positive. With the help of Eqs. (\ref{Prate}%
) and (\ref{rate}) one can calculate the nonlinear inverse-bremsstrahlung
absorption rate for Maxwellian, as well as for degenerate quantum plasmas.

\section{Numerical Results and Discussion}

For the obtained absorption rate (\ref{Prate}) one need to concretize the
ionic potential $V\left( \mathbf{q}_{s}\right) $. For s single ion of charge
number $Z_{a}$ we will assume screening Coulomb potential with radius of
screening $\varkappa _{e}^{-1}\ $as a function of the plasma temperature and
density of electrons $n_{e}$: 
\end{subequations}
\begin{equation*}
\varkappa _{e}=\left( 4\pi e^{2}\frac{\partial n_{e}}{\partial \mu }\right)
^{1/2}.
\end{equation*}%
Thus, taking into account the plasma quasi-neutrality ($Z_{a}N_{i}=N_{e}$),
we have 
\begin{equation}
\left\vert V\left( \mathbf{q}_{s}\right) \right\vert ^{2}=N_{e}\frac{16\pi
^{2}Z_{a}e^{4}}{\left( \mathbf{q}_{s}^{2}+\varkappa _{e}^{2}\right) ^{2}}.
\label{Vqq}
\end{equation}%
Integrating in Eq. (\ref{Prate}) over $E_{\mathbf{\Pi }^{\prime }}$ we will
obtain the following expression for partial absorption rates: 
\begin{equation*}
\frac{dN_{\gamma e}\left( s\right) }{dt}=\frac{2Z_{a}e^{4}s}{\pi ^{3}\hbar
^{3}c^{4}}\int_{m^{\ast }c^{2}+s\hbar \omega }dE_{\mathbf{\Pi }}d\Omega
d\Omega ^{\prime }\frac{\left\vert \mathbf{\Pi }\right\vert \left\vert 
\mathbf{\Pi }^{\prime }\right\vert \left\vert \mathcal{B}_{\mathbf{\Pi }%
^{\prime };\mathbf{\Pi }}^{(s)}\right\vert ^{2}}{\left( \hbar ^{2}\mathbf{q}%
_{s}^{2}+\hbar ^{2}\varkappa _{e}^{2}\right) ^{2}}
\end{equation*}%
\begin{equation}
\times \left( 1-e^{-\frac{s\hbar \omega }{T_{e}}}\right) n\left( E_{\mathbf{%
\Pi }}-s\hbar \omega \right) \left( 1-n\left( E_{\mathbf{\Pi }}\right)
\right) ,  \label{dNje}
\end{equation}%
where 
\begin{equation*}
\left\vert \mathbf{\Pi }^{\prime }\right\vert =\sqrt{\left\vert \mathbf{\Pi }%
\right\vert ^{2}+\hbar ^{2}s^{2}\mathbf{k}^{2}-2\frac{E_{\mathbf{\Pi }%
}s\hbar \omega }{c^{2}}}.
\end{equation*}

\begin{figure}[tbp]
\includegraphics[width=.5\textwidth]{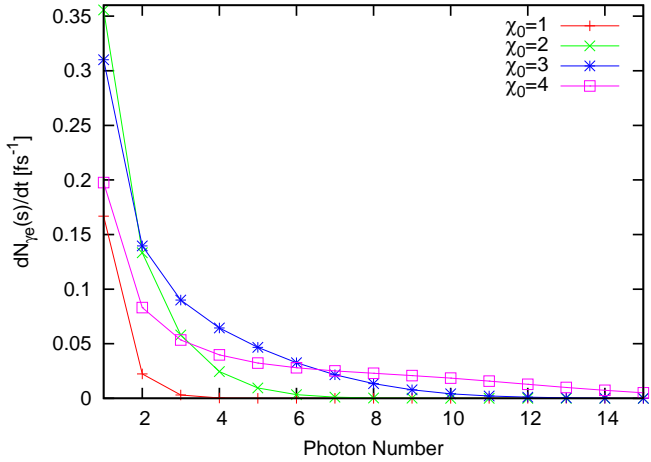}
\caption{(Color online) Envelope of partial rate of inverse-bremsstrahlung
absorption via the mean number of absorbed photons by per electron, per
unite time (in femtosecond$^{-1}$) for CPW in Maxwellian plasma is shown for
various wave intensities at $\protect\varepsilon _{\protect\gamma }=1\ 
\mathrm{keV}$, $Z_{a}=13$, $n_{e}=10^{23}\mathrm{cm}^{-3}$, and $T_{e}=100$ $%
\mathrm{eV}$. }
\end{figure}
\begin{figure}[tbp]
\includegraphics[width=.5\textwidth]{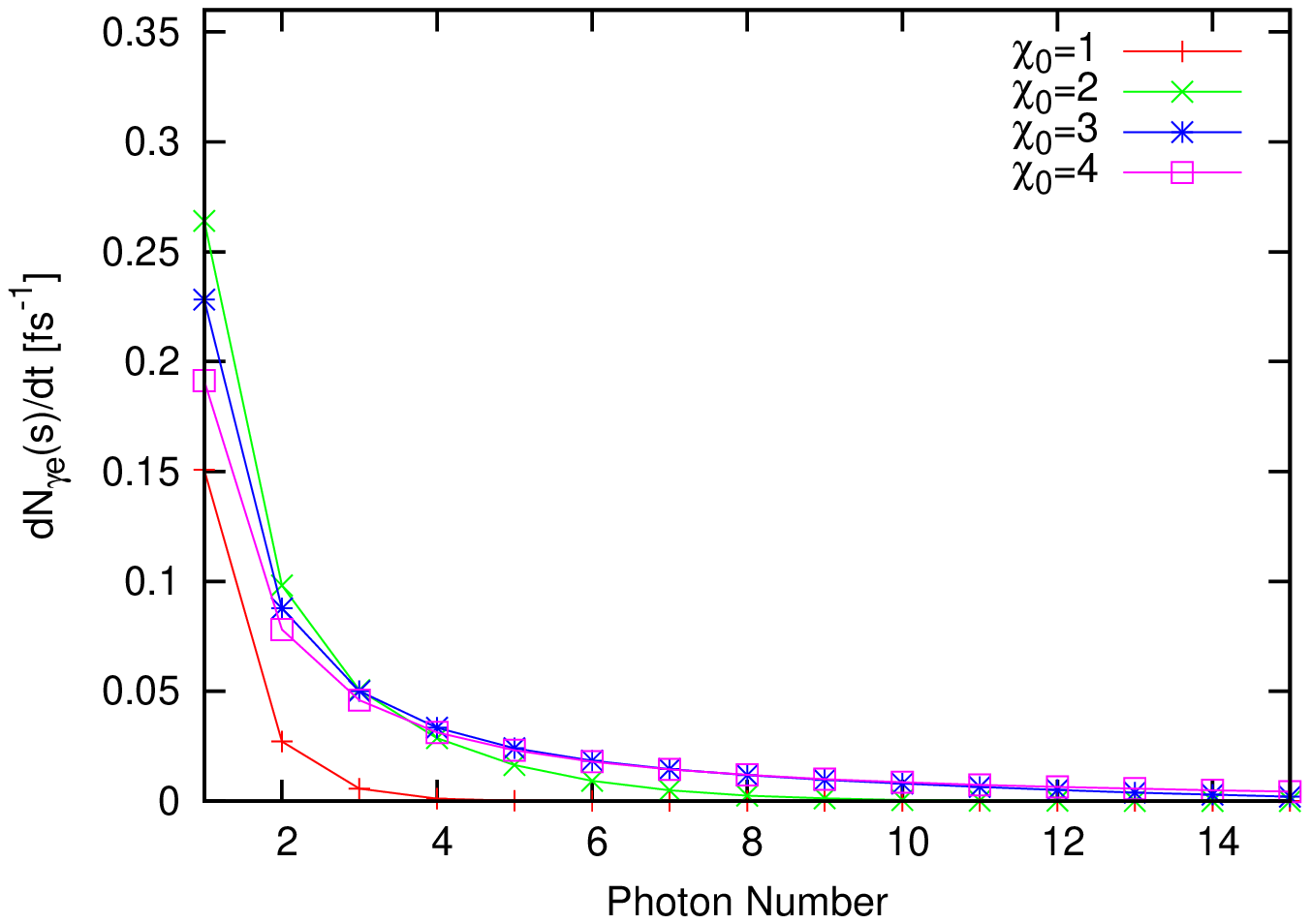}
\caption{(Color online) Same as figure 1 but for LPW. }
\end{figure}

\begin{figure}[tbp]
\includegraphics[width=.5\textwidth]{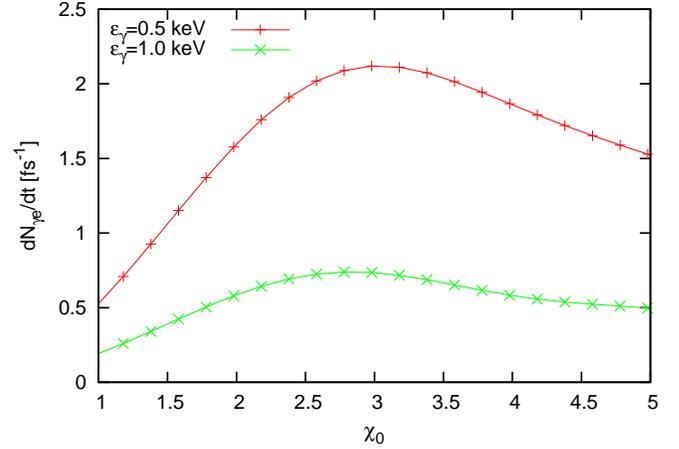}
\caption{(Color online) Total rate of inverse-bremsstrahlung absorption for
CPW in Maxwellian plasma versus the dimensionless parameter $\protect\chi %
_{0}$ for various photon energies at $n_{e}=10^{23}\mathrm{cm}^{-3}$, and $%
T_{e}=100$\ $\mathrm{eV}$. }
\label{098}
\end{figure}
\begin{figure}[tbp]
\includegraphics[width=.5\textwidth]{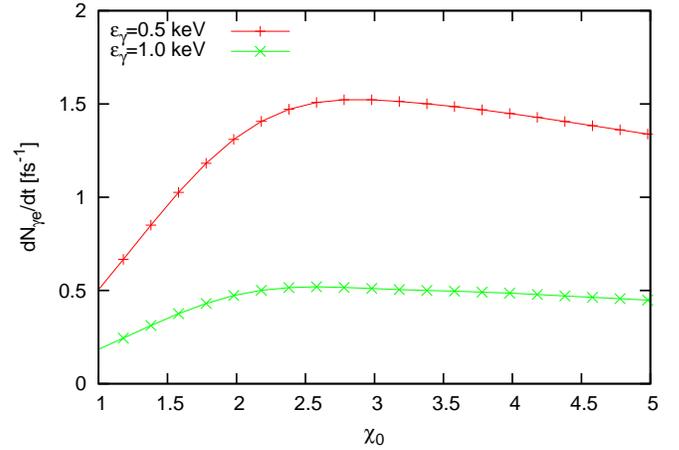}
\caption{(Color online) Same as figure 3 but for LPW. }
\end{figure}

In general, the analytical integration over solid angles $\Omega $, $\Omega
^{\prime }$ and energy is impossible, and one should make numerical
integration. The latter for initially nonrelativistic plasma and at the
photon energies $\hbar \omega >$ $T_{e}$, is convenient to made introducing
a dimensionless parameter 
\begin{equation}
\chi _{0}=\frac{eE_{0}}{\omega \sqrt{m\hbar \omega }},  \label{kappa}
\end{equation}%
which is the ratio of the amplitude of the momentum transferred by the wave
field to the momentum at the one-photon absorption. In Eq. (\ref{kappa}) the
dimensionless parameter $E_{0}=\omega A_{0}\sqrt{1+g^{2}}/c$ is the
amplitude of the electric field strength. Hence the average intensity of the
wave expressed via the parameter $\chi _{0}$, can be estimated as%
\begin{equation*}
I_{\chi _{0}}=\chi _{0}^{2}\times 1.74\times 10^{12}\ \mathrm{W\ cm}^{-2}%
\left[ \frac{\hbar \omega }{\mathrm{eV}}\right] ^{3}.
\end{equation*}%
The intensity $I_{\chi _{0}}$ strongly depends on the photon energy $\hbar
\omega $. At $\chi _{0}\sim 1$, the multiphoton effects become essential.
Particularly, for x-ray photons with energies $\varepsilon _{\gamma }\equiv
\hbar \omega =0.1-1\ \mathrm{keV}$, multiphoton interaction regime can be
achieved at the intensities $I_{\chi _{0}}\sim 10^{18}-10^{21}\ \mathrm{W/cm}%
^{2}$. In the opposite limit $\chi _{0}\ll 1$, the multiphoton effects are
suppressed.

For all calculations as a reference sample we take ions with $Z_{a}=13$
(fully ionized Aluminum) and consider plasma of solid densities. In Fig. 1
and Fig. 2 the envelope of the partial rate of inverse-bremsstrahlung
absorption for CPW and LPW in Maxwellian plasma is shown for various wave
intensities at $\varepsilon _{\gamma }=1\ \mathrm{keV}$, $n_{e}=10^{23}%
\mathrm{cm}^{-3}$, and $T_{e}=100$ $\mathrm{eV}$ ($n_{e}\propto e^{\mu
/T_{e}}$ and $\varkappa _{e}=\left( 4\pi e^{2}n_{e}/T_{e}\right) ^{1/2}$).
As is seen from this figures, the multiphoton effects become essential with
the increase of the wave intensity.

To show the dependence of the inverse-bremsstrahlung absorption rate on the
laser radiation intensity, in Fig. 3 and Fig. 4 the total rate (\ref{rate})
with (\ref{dNje}) for CPW and LPW in Maxwellian plasma versus the parameter $%
\chi _{0}$ for various photon energies are shown. As is seen from these
figures, the rate strictly depends on the wave polarization, and for the
large values of $\chi _{0}$ it exhibits a tenuous dependence on the wave
intensity.

To compare with\ the linear theory \cite{Bunk}, in Fig. 5 we plot scaled
absorption rate $\chi _{0}^{-2}dN_{\gamma e}/dt$ versus $\chi _{0}$. In the
scope of the linear theory the scaled absorption rate does not depend on the
wave intensity, while for the large values of $\chi _{0}$ it is suppressed
with the increase of the wave intensity. 
\begin{figure}[tbp]
\includegraphics[width=.5\textwidth]{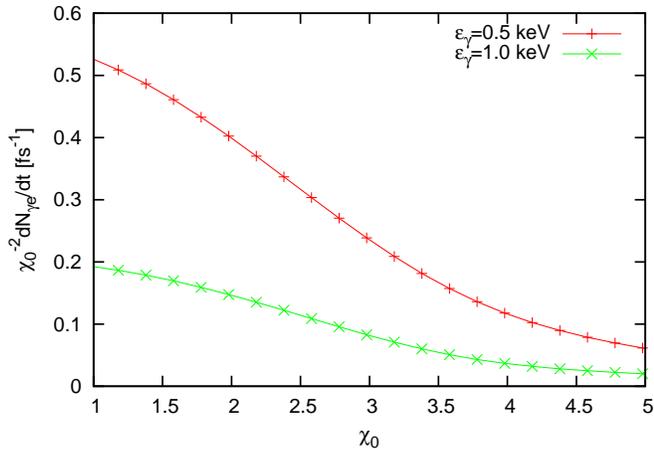}
\caption{(Color online) Total rates of the inverse-bremsstrahlung absorption
scaled to $\protect\chi _{0}^{2}$ versus the parameter $\protect\chi _{0}$
for setup of Fig 3. }
\end{figure}
To show the dependence of the considered process on the plasma temperature,
in Fig. 6 we plot total rate of the inverse-bremsstrahlung absorption of
circularly polarized laser radiation in Maxwellian plasma, as a function of
the plasma temperature for various wave intensities at $\varepsilon _{\gamma
}=1\ \mathrm{keV}$, and $n_{e}=10^{23}\mathrm{cm}^{-3}$. The similar picture
holds for LPW. Here for the large values of $\chi _{0}$ we have a weak
dependence on the temperature, which is a result of the laser modified
scattering of electrons irrespective of its' initial state. 
\begin{figure}[tbp]
\includegraphics[width=.5\textwidth]{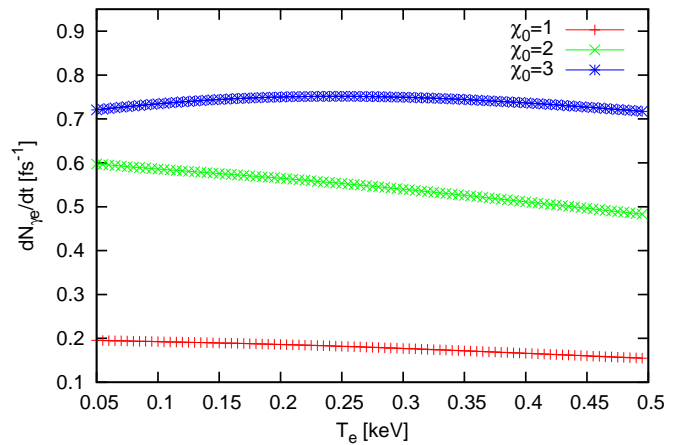}
\caption{(Color online) Total rate of inverse-bremsstrahlung absorption of
CPW in Maxwellian plasma, as a function of the plasma temperature is shown
for various wave intensities at $\protect\varepsilon _{\protect\gamma }=1\ 
\mathrm{keV}$, $Z_{a}=13$, $n_{e}=10^{23}\mathrm{cm}^{-3}$.}
\end{figure}
\begin{figure}[tbp]
\includegraphics[width=.5\textwidth]{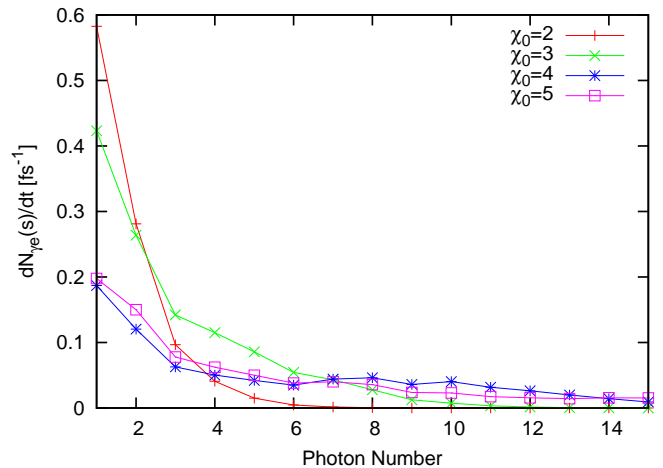}
\caption{(Color online) Envelope of partial rate of inverse-bremsstrahlung
absorption for CPW in degenerate plasma with Fermi energy $\protect%
\varepsilon _{F}=11.7$ $\mathrm{eV}$ is shown for various wave intensities
at $\protect\varepsilon _{\protect\gamma }=1\ \mathrm{keV}$, $Z_{a}=13$, and 
$T_{e}=0.1\protect\varepsilon _{F}$. }
\end{figure}
\begin{figure}[tbp]
\includegraphics[width=.5\textwidth]{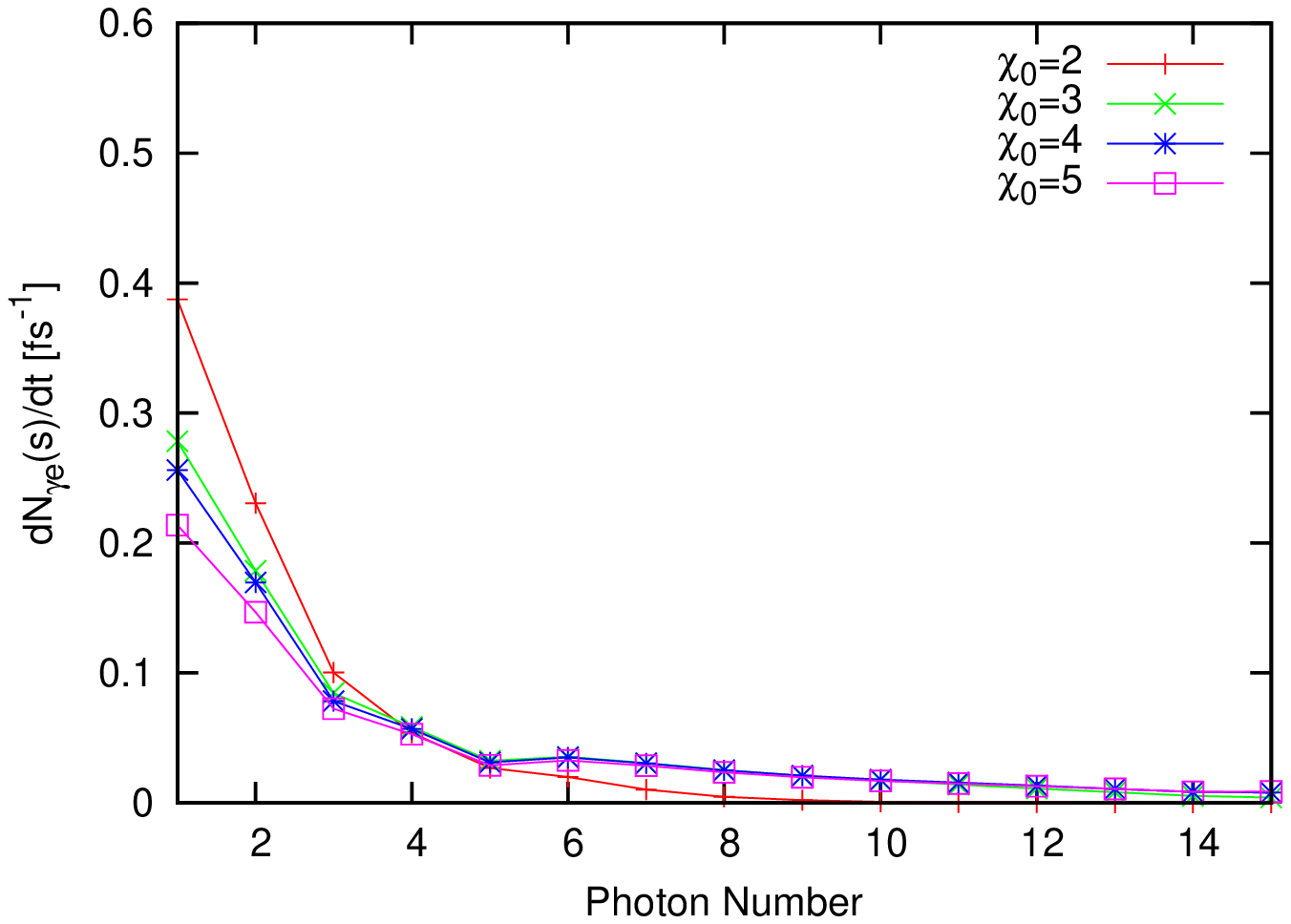}
\caption{(Color online) Same as figure 7 but for LPW. }
\end{figure}
\begin{figure}[tbp]
\includegraphics[width=.5\textwidth]{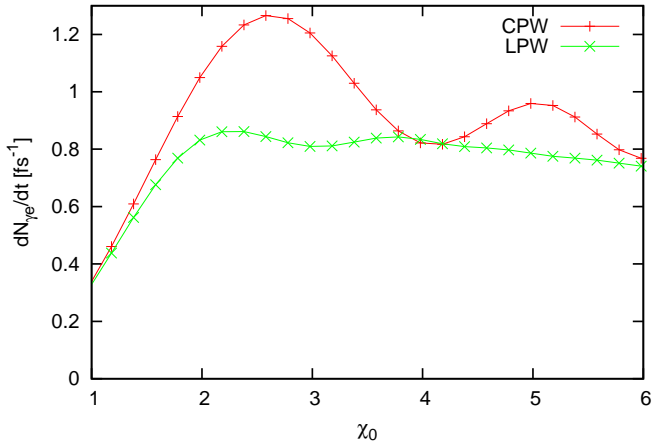}
\caption{(Color online) Total rate of inverse-bremsstrahlung absorption for
CPW and LPW in degenerate plasma versus the parameter $\protect\chi _{0}$ at 
$\protect\varepsilon _{F}=11.7\ \mathrm{eV}$, $Z_{a}=13$, and $T_{e}=0.1%
\protect\varepsilon _{F}$. }
\end{figure}
We have also made calculations for a degenerate quantum plasma. In Fig. 7
and Fig. 8 the envelope of the partial rate of inverse-bremsstrahlung
absorption for CPW and LPW in degenerate plasma with Fermi energy $\mu
\simeq \varepsilon _{F}=11.7$ $\mathrm{eV}$ (Aluminum) is shown for various
wave intensities at $\varepsilon _{\gamma }=1\ \mathrm{keV}$, and $%
T_{e}=0.1\varepsilon _{F}$ ($\varkappa _{e}=\left( 6\pi
e^{2}n_{e}/\varepsilon _{F}\right) ^{1/2}$). The total rate of
inverse-bremsstrahlung absorption via the mean number of absorbed photons by
per electron, per unite time for CPW and LPW in degenerate plasma versus the
parameter $\chi _{0}$ at $\varepsilon _{F}=11.7\ \mathrm{eV}$ and $%
T_{e}=0.1\varepsilon _{F}$ is shown in Fig. 9. As is seen from these
figures, the rate strictly depends on the wave polarization, and for the
large values of $\chi _{0}$ it is saturated. Note that our consideration is
valid when the pulse duration $\tau $ of an EM wave is restricted by the
condition 
\begin{equation}
\tau <\nu _{eff}^{-1},  \label{bc}
\end{equation}%
where $\nu _{eff}^{-1}$ is the time scale during which the thermalization of
the electrons energy in plasma occurs. In the presence of a laser field the
electron-ion binary collisions take place with the effective frequency 
\begin{equation*}
\nu _{eff}\simeq \frac{2\pi Z_{a}e^{4}n_{e}}{m^{2}\left\langle \mathrm{v}%
\right\rangle ^{3}}L_{\mathrm{cb}},
\end{equation*}%
where $L_{\mathrm{cb}}$ is the Coulomb logarithm, and $\left\langle \mathrm{v%
}\right\rangle $ is the mean values of electrons velocity in the laser
field. For moderate intensities one can write $\left\langle \mathrm{v}%
\right\rangle \simeq \chi _{0}\sqrt{\varepsilon _{\gamma }/m}$. For the
considered parameters we have $\nu _{eff}\simeq 10^{14}-10^{15}\mathrm{s}%
^{-1}$. Thus, the pulse duration should be $\tau <1\ \mathrm{fs}$. The
latter is satisfied for x-ray sources. As is seen from Figs. 3, 4, and 9,
for the pulse durations $\tau \lesssim 1\ \mathrm{fs}$ one can achieve an
one absorbed x-ray photon by per electron, which means that in plasma of
solid densities one can reach the plasma heating of high temperatures by
x-ray laser with the intensity parameter $\xi \sim 0.1$ already at the one
photon absorption process.

\section{Conclusion}

Concluding, we have presented the microscopic relativistic quantum theory of
multiphoton inverse-bremsstrahlung absorption of an intense shortwave laser
radiation in the classical and quantum plasma. The Liouville-von Neumann
equation for the density matrix has been solved analytically considering a
wave-field exactly, while a scattering potential of the plasma ions as a
perturbation. With the help of this solution we derived a relatively compact
expression for the nonlinear inverse-bremsstrahlung absorption rate when
electrons are \ represented by the grand canonical ensemble. Numerical
investigation of the obtained results for Maxwellian, as well as degenerate
quantum plasmas at x-ray frequencies and large values of laser fields has
been performed. The obtained results demonstrate that for the shortwave
radiation the SB rate being practically independent of the plasma
temperature is saturated with the increase of the wave intensity. The
obtained results showed that in the x-ray domain of frequencies one can
achieve an efficient absorption of powerful radiation specifically in plasma
of solid densities and reach the plasma heating of high temperatures.

\begin{acknowledgments}
This work was supported by SCS of RA under Project No. 13-1C066.
\end{acknowledgments}

\end{document}